\documentclass[aip,twocolumn,a4paper,superscriptaddress,floatfix,longbibliography,showkeys]{revtex4}
\usepackage{graphicx,amsfonts,amsmath,bm,color,tabularx}
\usepackage{multirow,hyperref}
\usepackage[normalem]{ulem}

\makeatletter
\let\@fnsymbol\@fnsymbol@latex
\@booleanfalse\altaffilletter@sw
\makeatother

\begin{document}

\title{Anomalous lattice specific heat and rattling phonon modes in quadruple perovskites}

\author{Valentin Yu. Irkhin}
\affiliation{Institute of Metal Physics, S. Kovalevskaya str. 18, 620108 Ekaterinburg, Russia}

\author{Zhehong Liu}
\affiliation{Beijing National Laboratory for Condensed Matter Physics, Institute of Physics, Chinese Academy of Sciences, Beijing 100190, China}

\author{Danil A. Myakotnikov}
\affiliation{Institute of Metal Physics, S. Kovalevskaya str. 18, 620108 Ekaterinburg, Russia}

\author{Evgenia V. Komleva}
\affiliation{Institute of Metal Physics, S. Kovalevskaya str. 18, 620108 Ekaterinburg, Russia}

\author{Youwen Long}
\affiliation{Beijing National Laboratory for Condensed Matter Physics, Institute of Physics, Chinese Academy of Sciences, Beijing 100190, China}

\author{Sergey V. Streltsov}
\affiliation{Institute of Metal Physics, S. Kovalevskaya str. 18, 620108 Ekaterinburg, Russia}

\date{\today}

\date{\today}
\begin{abstract}
Experimental data on the specific heat  $C_p$  of quadruple perovskites \textit{A}Cu$_3$Fe$_2$Re$_2$O$_{12}$ (\textit{A} = Mn, Cu, La, Ce, Dy) are presented, demonstrating an anomalous concave-down  $C_p/T$  vs.  $T^2$  curve and a bell-shaped feature in  $\beta(T) = (C_p - \gamma T)/T^3$  plotted against $T$ on a logarithmic scale. This feature is most pronounced for \textit{A} = Cu and Mn. These findings can be explained by the rattling phenomenon, previously identified in other systems such as filled skutterudites and $\beta$-pyrochlores.  Using first-principles DFT+U calculations, the presence of a rattling mode in \textit{A }= Mn is directly confirmed.  A qualitative interpretation of the rattling mechanism in terms of a pseudo-Jahn-Teller effect is proposed.
\end{abstract}
\maketitle

\section{Introduction}

Strong anharmonicity of the lattice potential, including its double-well or  multi-well behavior, 
and the presence of the localized (so-called rattling) phonon modes can result in a number of non-trivial effects in electronic and lattice properties.
 Such effects in quasi-two-dimensional cuprates, three-dimensional perovskites, filled skutterudites, $\beta$-pyrochlore oxides and related systems are extensively  discussed \cite{IKT,Menushenkov,Akizuki2015,Matsuhira,Keppens,Yamaura2006,Sc6MTe2}.

As confirmed  by many experimental facts, oxygen atoms at the corners of the CuO$_6$ octahedra in cupric oxides are in a double-well potential. 
In Ref.\cite{Bishop} the appearance of the double-well potential in superconducting La$_{2}$CuO$_4$ was explained in terms of the Jahn-Teller polaron model. Calculations of zone-center soft modes,  carried out to characterize the polar and octahedral-rotation instabilities  \cite{Roy}, confirmed the existence of a double-well potential in double perovskites.
The unique oxygen vibrations within a double-well potential are a characteristic commonly found in superconducting oxides possessing perovskite-like structures \cite{Menushenkov}.

{ Vibrations of weakly bound ions within a spacious atomic cage created by surrounding atoms, which correspond to large lattice distortion and become anharmonic, are commonly referred to as rattling.} This phenomenon has been observed in a number of materials such as
VAl$_{10+\delta}$~\cite{Caplin1973}, clathrates~\cite{Eisenmann1986}, dodecaborides~\cite{Sluchanko2018}, filled skutterudites~\cite{Braun1980,Matsuhira,Keppens}, $\beta$-pyrochlore oxides~\cite{Yamaura2006}. 
{ 
Rattling phenomenon was also described in the binary InTe \cite{Jana2016},
Cu$_{12}$Sb$_4$S$_{13}$ tetrahedrites with promising thermoelectric properties \cite{Lara,Lai}, molybdenum selenides containing Mo clusters Ag$_x$Mo$_9$Se$_{11}$ \cite{Zhou}.
Some of these systems demonstrate rattling Einstein modes with rather high frequencies, which is confirmed by concave dependences of specific heat with pronounced maxima.

}

Rattling or other forms of anharmonicity can also result in a Schottky-type anomaly in the specific heat $C_p$ at low temperatures \cite{Hasegawa}, lead to the significant increase of electron effective mass \cite{Brihwiler2006,Grosche2001,Bauer2002}, suppress thermal conductivity~\cite{Jana2016,Chang2018} or be a motivating factor for the superconductivity~\cite{Brihwiler2006, Grosche2001, Bauer2002, Nagao2009}.

In $\beta$-pyrochlore oxides KOs$_2$O$_6$ and RbOs$_2$O$_6$,
an unusual energetically low-lying phonon is connected with a rattling motion of the alkali-metal ion \cite{Brihwiler2006}. 
The rattling is a general property for filled skutterudites La$T_4X_{12}$ (\textit{T} = Fe, Ru, and Os; \textit{X} = P, As, and Sb) where the existence of low-energy guest-ion optical modes (LGOMs) was proposed \cite{Matsuhira}. Lattice vibrations result in the famous Debye law, when specific heat is proportional to $T^3$, while the electronic subsystem contributes linearly with $T$ in the Fermi liquid regime~\cite{Kittel-book}. Therefore, naively one might expect that $(C_p-\gamma T)/T^3$ would not depend on temperature, but in LGOMs a broad maximum in $(C_p-\gamma T)/T^3$ is observed at low temperatures, $T_{max} = 10-30$ K. This value is ascribed to $\Theta_E/5$, where $\Theta_E$ is the Einstein temperature \cite{Matsuhira}. The NdOs$_4$Sb$_{12}$ crystal can be considered to be a Debye solid, but with the Nd atoms acting like Einstein oscillators, $\Theta_E$ estimated to be about 45 K. This value is close to that found for thallium-filled antimony skutterudites such as Tl$_{0.22}$Co$_4$Sb$_{12}$ \cite{Ho}.
 Rattling of R ions in  ROs$_4$Sb$_{12}$ was confirmed in EXAFS spectroscopy by observation of the static distortions of the interatomic separation of the R–Sb atomic pairs,  i.e., of the local potential minimums for the R ions \cite{Nitta1}. 
\begin{figure}[b!]
\includegraphics[width=0.46\textwidth]{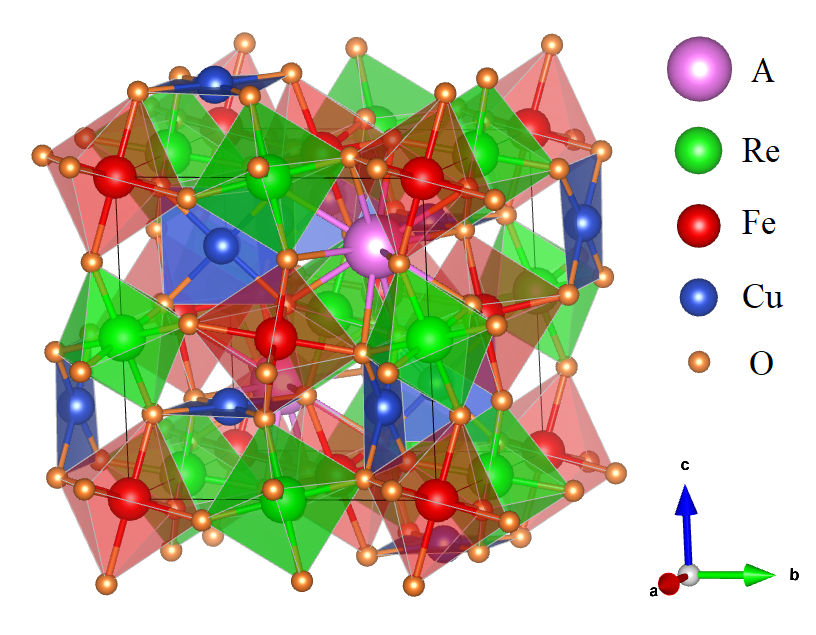}
\caption{\label{fig:crystal-structure} { The crystal structure of \textit{A}Cu$_3$Fe$_2$Re$_2$O$_{12}$ quadruple perovskites. $A$ ions sits in the distorted icosahedral cage and can rattle.}}
\end{figure}

More recently,  rattling has been suggested for the  quadruple perovskite CuCu$_3$V$_4$O$_{12}$~\cite{Akizuki2015}. In quadruple perovskites $AA'_3$B$_4$O$_{12}$ the $A$ site ions are icosahedrally (with twelve neighbors) coordinated by oxygen atoms. The thermal displacement parameter of Cu ions at $A$ site in CuCu$_3$V$_4$O$_{12}$ was found to be rather large, $U_{iso}$$\approx$0.045~\AA$^2$ at 300 K. In contrast with CaCu$_3$V$_4$O$_{12}$, CuCu$_3$V$_4$O$_{12}$ exhibits a concave-down $C_p/T$ vs. $T^2$
curve, which immediately suggests a low-frequency enhancement
of vibrational density of states \cite{Akizuki2015}.
The dominance of local modes is also confirmed by a bell-shaped behavior of $(C_p-\gamma T)/T^3$ vs. $T$ on a logarithmic scale of around 10 K.  Such behavior cannot be described by a Debye-type phonon model but can be pronounced in the presence of rattling. Thus, the unusual behavior of the specific heat $C_p(T)$ leads to the conclusion that rattling occurs in CuCu$_3$V$_4$O$_{12}$.

{ Theoretical calculations  revealed that rattling behavior is present in both in CuCu$_3$V$_4$O$_{12}$ and CuCu$_3$Fe$_2$Re$_2$O$_{12}$ \cite{Pchelkina}. They clearly show rattling distortions along [111] direction  related to the Cu vibration. Moreover, there are local minima not only in this direction, but also in [001] and [110] ones. It was shown that Cu ions at $A$ sites vibrate in the center of the icosahedral oxygen O$_{12}$ cages and the corresponding potential is not of a simple double-well form, but has a complicated form with many local minima. It is of interest to confirm this theoretical finding by experimental observations.}

In the present paper,  new experimental results are presented that reveal an anomalous temperature dependence of the specific heat for a number of quadruple perovskites in the \textit{A}Cu$_3$Fe$_2$Re$_2$O$_{12}$ series. These compounds demonstrate interesting half-metallic ferromagnetic properties that were previously investigated both experimentally and theoretically for \textit{A} = Cu \cite{Pchelkina,Cu}, Ca \cite{Chen,Wang1,Zhang1,Zhang,Miao},  La \cite{Zhang,Liu1,Wang1}, Ce \cite{Peng}, Mn \cite{MCFRO_Crystal_structure}, Dy \cite{Dy}; { the corresponding first-principle calculations of magnetic configurations and exchange interactions were performed in Ref. \cite{magn}.} 
Here, we analyze the data on specific heat and extract their characteristic parameters. The experimental findings are supplemented by theoretical calculations, which explicitly demonstrate rattling in the two quadruple perovskites exhibiting anomalous temperature dependence -- MnCu$_3$Fe$_2$Re$_2$O$_{12}$ and CuCu$_3$Fe$_2$Re$_2$O$_{12}$.
\begin{figure}[t!]
\centering
\includegraphics[width=1\columnwidth]{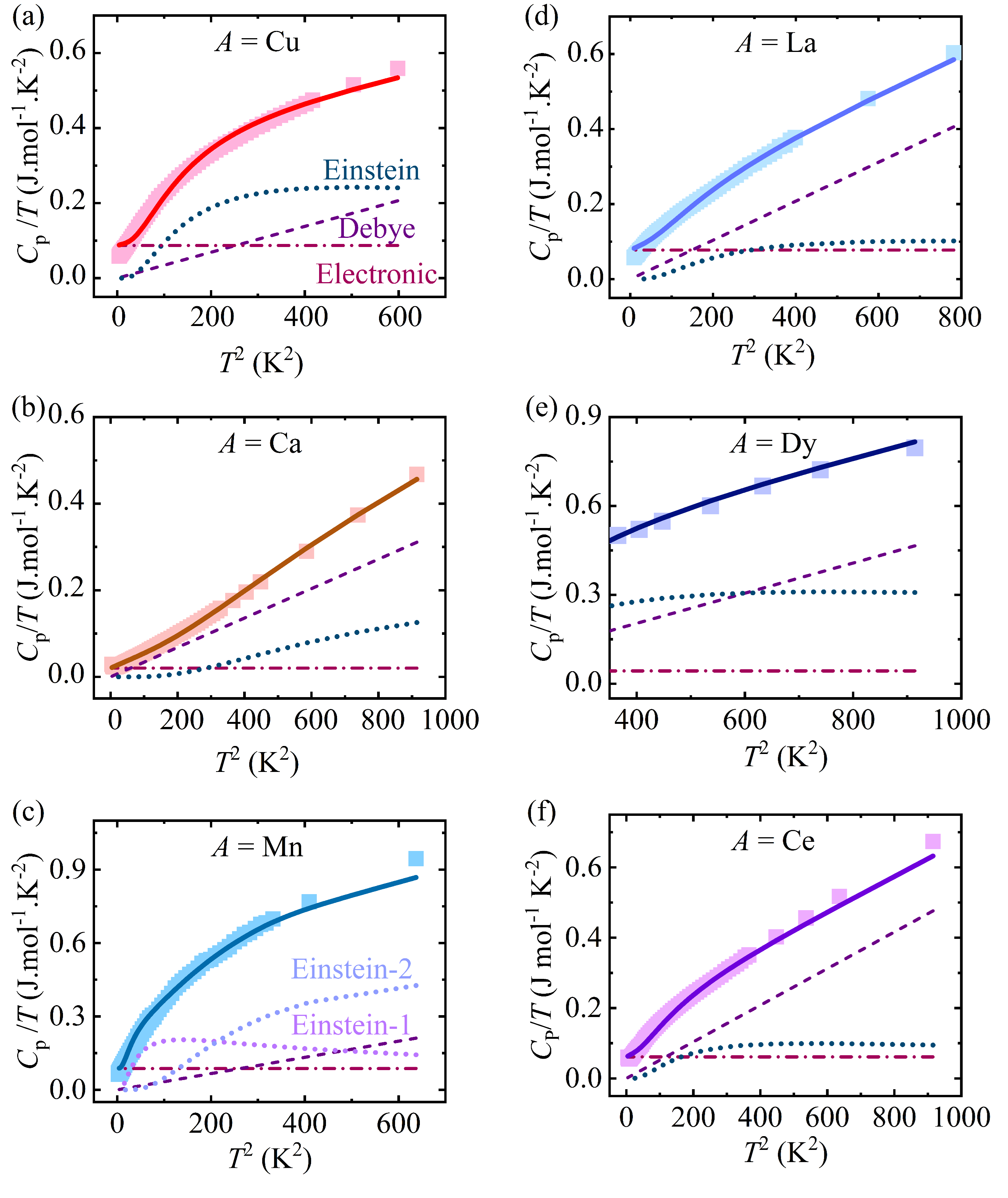}
\caption{$C_p/T$ vs $T^2$ of quadruple perovskites $A$Cu$_3$Fe$_2$Re$_2$O$_{12}$, (a) $A$ = Cu, (b) $A$ = Ca, (c) $A$ = Mn, (d) $A$ = La, (e) $A$ = Dy, $A$ = Ce. The solid lines show the fitting results basing on function (\ref{C_p_fitting}) with one $r$ parameter for $A$Cu$_3$Fe$_2$Re$_2$O$_{12}$ ($A$=Cu, Ca, La, Ce and Dy) and two parameters $r_1$ and $r_2$ in the sum for MnCu$_3$Fe$_2$Re$_2$O$_{12}$. { The Sommerfeld (electronic), Debye and Einstein contributions are shown by different lines, notations being given in  (a).}}
	\label{F_1}
\end{figure}


\section{\label{exp} Experimental}

{ 

Polycrystalline samples of Cu/MnCu$_3$Fe$_2$Re$_2$O$_{12}$  were synthesized from stoichiometric mixtures of high-purity ($>$99.9~\%) powders of oxides MnO, CuO, Fe$_2$O$_3$ and Re$_2$O$_7$, and of Re. The thoroughly ground mixtures were pressed into gold or platinum capsules and subjected to a pressure of 18~GPa, followed by heating at 1373 K for one hour. After heating, the samples were quenched to room temperature and the pressure was gradually released. A similar procedure was employed for the synthesis of Ce/Ca/DyCu$_3$Fe$_2$Re$_2$O$_{12}$  compounds, except that a lower pressure of 8 GPa was applied.

Powder X-ray diffraction (XRD) measurements were performed using a Huber diffractometer with Cu K$\alpha$ radiation ($\lambda$ = 1.5406\textup{~\AA}). Specific heat  measurements were carried out using the pulse relaxation method on a Physical Property Measurement System (PPMS). The temperature range was 2--200 K at zero field for Cu/Mn/Ce/CaCu$_3$Fe$_2$Re$_2$O$_{12}$  samples, and at magnetic fields of 0, 0.1, 0.3, 1.0, and 3.0 T for DyCu$_3$Fe$_2$Re$_2$O$_{12}$.

The crystal structure of the \textit{A}Cu$_3$Fe$_2$Re$_2$O$_{12}$ quadruple perovskites is shown in Fig.~\ref{fig:crystal-structure}.
The detailed XRD data can be found for  \textit{A} = Cu in Ref. \cite{Cu}, for  \textit{A} = Ca  in Ref. \cite{Chen,Miao},  for  \textit{A} = La in Ref. \cite{Liu1}, for  \textit{A} = Ce in Ref. \cite{Peng}, for  \textit{A} = Mn in Ref. \cite{MCFRO_Crystal_structure}, for  \textit{A} = Dy in Ref. \cite{Dy}. }


\section{\label{exp-data} Results on specific heat}

The results for specific heat  of the quadruple perovskites \textit{A}Cu$_3$Fe$_2$Re$_2$O$_{12}$ (\textit{A} = Cu, Ca, Mn, La, Dy, Ce)    are presented in Fig.~\ref{F_1}. 
One can see that concave-down dependence of $C_p/T$ vs $T^2$ is most pronounced for $A$ = Cu and Mn. As discussed above, the anomalous behavior of non-linear contribution to specific heat means presence of anharmonic phonon modes.
Usually such a situation is treated in terms of Einstein-like low-energy phonons with a characteristic temperature $\Theta_E$ being much smaller (probably, by about two orders of magnitude) than the Debye temperature. Similar to Ref. \cite{Akizuki2015}, the following fitting procedure is  used. 

{ 
First, one has to set the values of  $\Theta_D$. A value standard for quadruple perovskites is about 500 K~\cite{Akizuki2015,Krimmel}. However, in systems containing one heavy
rare-earth atom per formula unit (containing many atoms), $\Theta_D$ should be somewhat lower. For  example, in \textit{A}Cu$_3$Ru$_4$O$_{12}$ the values of $\theta_D$, being near 500 K, are smaller for \textit{A} = La, Nd than for \textit{A} = Na, Ca by about 5-10\%  \cite{Gunther}. 
In our fitting, the values presented  in  Table \ref{Tab_1} turned out to be suitable.

Since we work at $T \ll  \Theta_D$, it is possible to use, as in Ref. \cite{Akizuki2015}, the simple $T^3$-approximation for the Debye lattice specific heat. 
Then, basing on the function which includes several possible Einstein modes with related temperatures $\Theta_{Ei}$  and weights $r_i<1$,
\begin{eqnarray}
\label{C_p_fitting}
C_p=\gamma T &+& \frac{12\pi^4}{5}\left(20-\sum_i r_i \right)R \frac{T^3}{\Theta_D^3} \nonumber \\
&+& \sum_i \frac{3r_iR({\Theta_{Ei}}/{T})^2\exp({\Theta_{Ei}}/{T})}{(\exp({\Theta_{Ei}}/{T})-1)^2},
\end{eqnarray}
($R$ is the gas constant) we get the fitted values of $\Theta_{Ei}$, $r_i$, and $\gamma$. The detailed results are shown in Table \ref{Tab_1}. 

{ Of course, such a fitting with many parameters cannot be fully perfect, but it should provide a qualitative picture.}
One can see large values of $\gamma $ for 
\textit{A} = Cu, Ce,  Mn, La, and especially Dy, but not for Ca.
A similar situation is observed in $A$Cu$_3$V$_4$O$_{12}$,  where
$\gamma$ = 126 mJ~mol$^{-1}$ K$^{-2}$ and 18-30 mJ~mol$^{-1}$ K$^{-2}$ for $A$ = Cu and $A$ = Ca (see  \cite{Akizuki2015}).
\begin{figure}[t!]
\centering
\includegraphics[width=1\columnwidth]{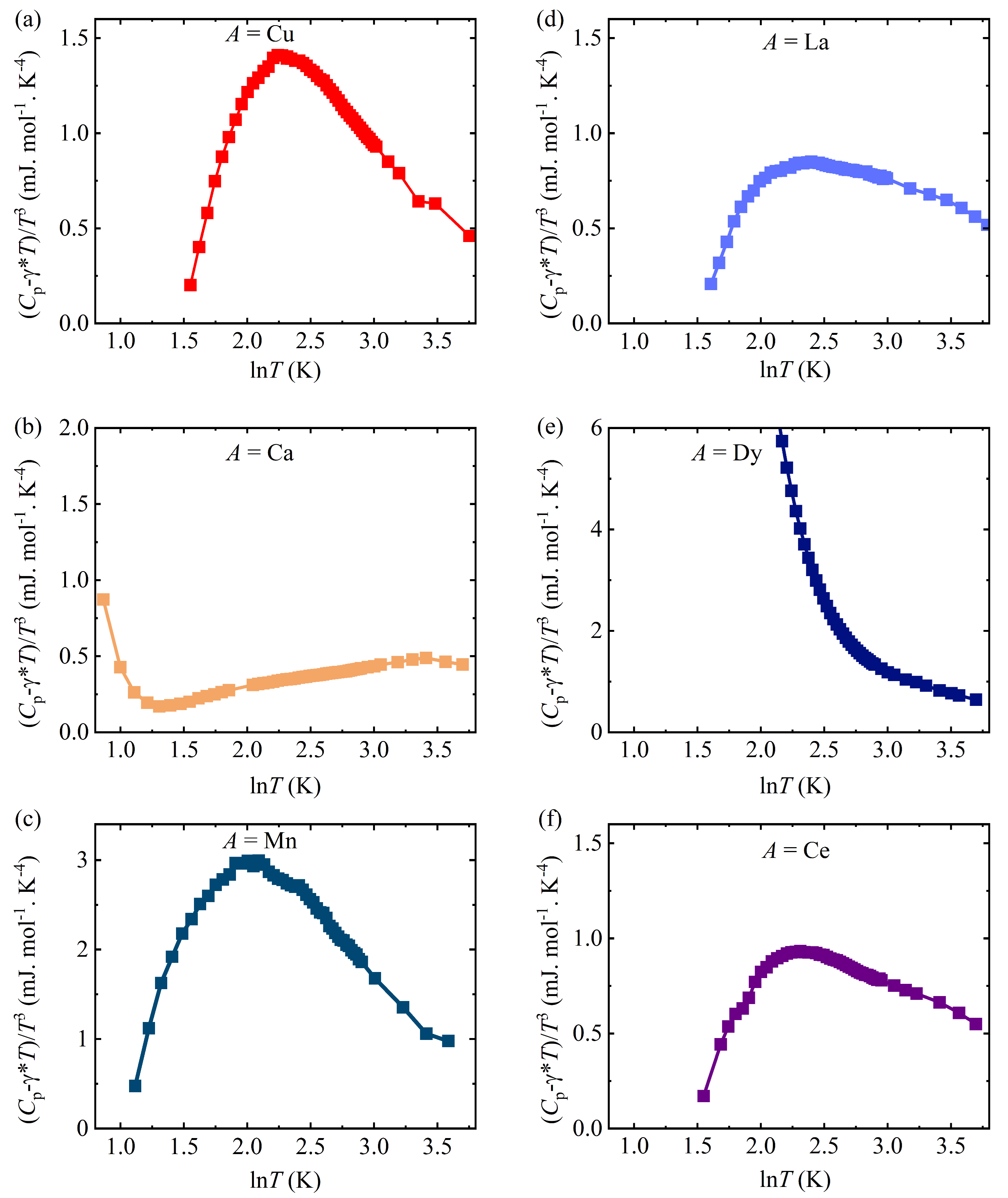}
\caption{{Temperature-dependent $\beta(T)=(C_p-\gamma T)/T^3$ of quadruple perovskites $A$Cu$_3$Fe$_2$Re$_2$O$_{12}$. (a) $A$ = Cu, (b) $A$ = Ca, (c) $A$ = Mn, (d) $A$ = La, (e) $A$ = Dy and (f) $A$ = Ce.}}
	\label{F_2}
\end{figure}

\begin{table}[b!]
 \begin{tabular}{ccccc}
 \hline
 \hline 
\textit{A}      & $\gamma$ (mJ/mol K$^{2}$) & $r$   & $\Theta_D$ (K)       & $\Theta_E$ (K)     \\
\hline
Cu & 87                           & 0.37     &480          &  58 \\
Ca & 20                             & 0.66     & 480           &  136\\
Mn & 87                     & 0.16($r_1$) & 480 & 30($\Theta_{E1}$)\\
                             &                            & 0.96($r_2$) & &81($\Theta_{E2}$) \\
La & 77                             & 0.19     & 420     &   71    \\
Dy & 43                            & 0.58     &425      &  71    \\
Ce & 66                            & 0.16     & 420       &  61   \\
 \hline
 \hline
 \end{tabular}
	\caption{{  The Einstein and Debye temperatures, and Sommerfeld coefficient $\gamma$ of the quadruple perovskites \textit{A}Cu$_3$Fe$_2$Re$_2$O$_{12}$} obtained by fitting Eq.~\eqref{C_p_fitting} with one parameter $r$ for $A=$ Cu, Ca, La, Ce and Dy and two parameters $r_1$ and $r_2$ in the sum for $A=$ Mn.}
		\label{Tab_1}
\end{table}

Using the fitted value of $\gamma$, we can plot the curves $\beta(T)=(C_p-\gamma T)/T^3$, as shown in Fig.~\ref{F_2}. They demonstrate some maxima in the temperature dependences.
As compared to CuCu$_3$V$_4$O$_{12}$ \cite{Akizuki2015}, we deal with lower energies of anharmonic phonon modes, so that the position of maximum in $\beta(T)$ is shifted to low $T_{max}$ about $10$ K, except for $A$ = Ca, Dy. Therefore, the bell-shaped form of $\beta(T)$ curve is not so clearly pronounced in the low-temperature limit, since measurements and separation of various contributions at very low $T$ are difficult. However, the behavior of lattice specific heat is clearly not described by the Debye law: { in the low temperature region}, the $\beta(T)$-term strongly decreases with increasing $T$, especially in Cu and Mn systems.

Whereas in most cases the description in terms of a single Einstein mode turns out to be sufficient,  the situation for MnCu$_3$Fe$_2$Re$_2$O$_{12}$ is more complicated. 
{ As well as other quadruple perovskites under consideration, this compound demonstrates ferrimagnetic ordering, $T_C$ being  about 400~K, as obtained experimentally from the magnetization curve \cite{MCFRO_Crystal_structure}.}
Here, we have fitted specific heat with two distinct Einstein temperatures.
A similar situation occurs in a number of other materials \cite{Gamsjaeger,Keppens,Gunther}, the corresponding frequencies, however, being as a rule considerably larger.
For example, in La-filled skutterudite  antimonide La$_{0.9}$Fe$_3$CoSb$_{12}$ contributions from two Einstein oscillators with two  Einstein temperatures $\Theta_{E1}$ = 70~K and $\Theta_{E2}$ = 200~K were introduced, the first one being ascribed to the rattling of the La ions oscillating in a harmonic potential \cite{Keppens}. 

\begin{figure}[t!]
\centering
\includegraphics[width=1\columnwidth]{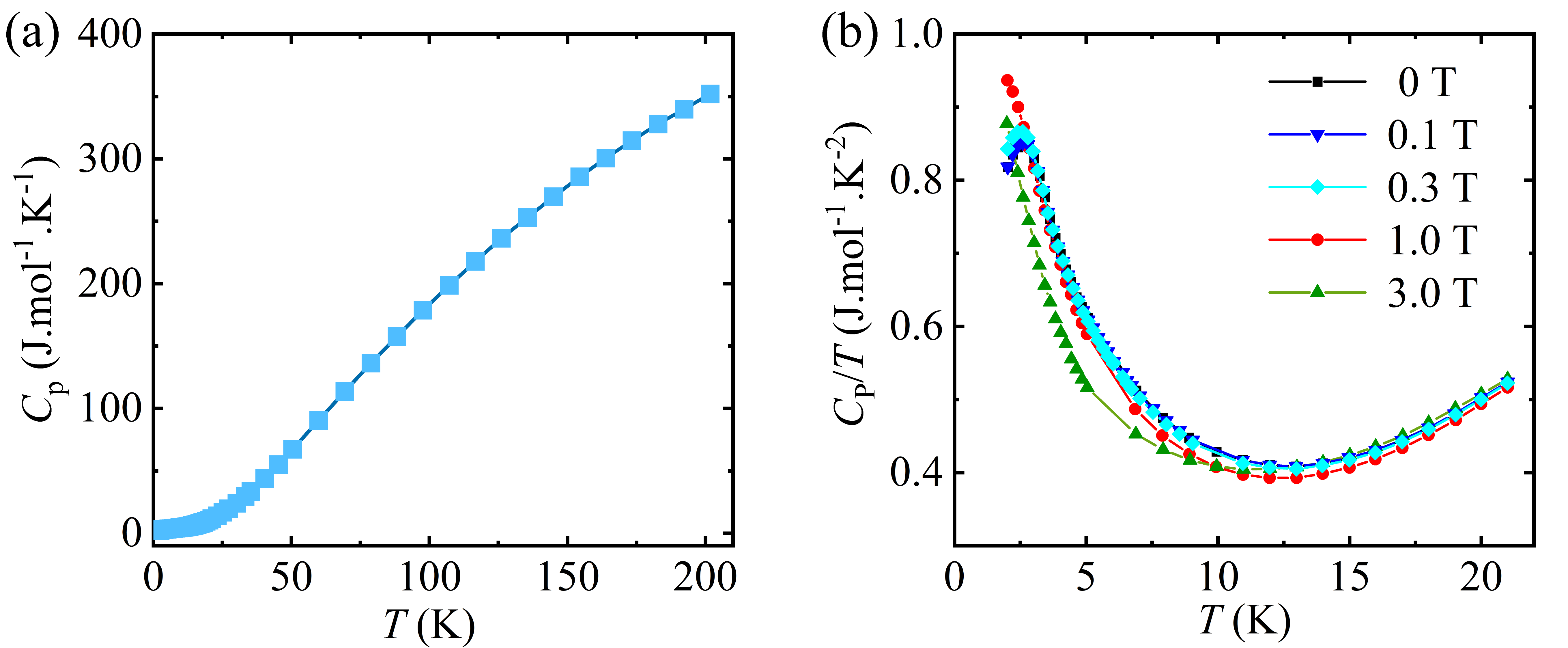}
\caption{{(a) Temperature dependence of specific heat of DyCu$_3$Fe$_2$Re$_2$O$_{12}$ between 2 and 201 K. (b) Temperature dependence of $C_p/T$ measured at some selected magnetic fields.}}
	\label{F_3}
\end{figure}



The situation in the Ca quadrupole perovskite is unclear: a broad maximum in $\beta(T)$ appears at higher temperatures. Therefore, if present, both Einstein-like and Debye contributions may play significant roles, with the Einstein mode frequency being considerably higher. This makes rattling for the Ca ion much less favorable. On the basis of the corresponding ionic radius (as discussed in the next section), it is expected that rattling is either suppressed or occurs with a much smaller amplitude. A somewhat similar difference is also observed between CuCu$_3$V$_4$O$_{12}$ and CaCu$_3$V$_4$O$_{12}$ \cite{Akizuki2015}.

 { For DyCu$_3$Fe$_2$Re$_2$O$_{12}$ \cite{Dy} the situation is  complex too because of possible influence of a  Schottky anomaly and antiferromagnetic ordering.
As shown in Fig.~\ref{F_3}, the temperature-dependent  $C_p/T$  exhibits an anomaly near 13~K which shifts to lower temperatures with increasing magnetic field. 
The situation is somewhat similar to Ref.~\cite{Adhikari} where the Schottky peaks in the filled skutterudite system Pr$_x$Eu$_{1-x}$Pt$_4$Ge$_{12}$ shift to higher temperatures with increase of magnetic field, signaling the presence of an internal magnetic field due to short-range AFM correlations induced by magnetic moments of neighboring Eu sites.
 According to \cite{Dy}, 
 the Dy-$M_{4,5}$ XMCD signal increases in intensity upon cooling when the temperature is above 100~K. However, below 100~K -- such as at 20 K -- the intensity drops sharply. This behavior suggests that the net magnetic moment of Dy$^{3+}$ first grows and then diminishes with decreasing temperature, providing clear evidence that an antiferromagnetic transition develops in the Dy sublattice at low temperatures. 
 
 Field and temperature dependences of magnetization have been also measured. It is clearly seen from Fig.~3b in \cite{Dy} that Dy spins at temperatures as low as 2~K are ordered and magnetic field can readily reorient them, while for $T =20$~K one cannot obtain fully polarized  magnetic state. This suggests that at low temperatures one has an ordered magnetic phase. Thus the anomaly in specific heat near 15~K can be considered as a feature corresponding to onset of AFM order. 
In the temperature range from 50 K down to about 15~K, short-range AFM correlations are most likely formed, making the precise determination of  $\gamma$  challenging. An attempt to fit  $C_p/T =\gamma + \beta(T)T^2$  at intermediate temperatures results in a strongly temperature-dependent  $\beta(T)$. An estimate considering a single Einstein mode yields a large value of $\gamma$, approximately 45 mJ~mol$^{-1}$ K$^{-2}$.
}

{ 
Since the quadruple perovskites are ferromagnetic, a question arises whether the specific heat can be influenced by the magnon $T^{3/2}$-contribution.
The problem of extracting this contribution is not simple and different results are obtained for different ferromagnets. In metallic manganites La$_{0.67}$ Ca$_{0.33}$MnO$_{3-\delta}$  (where the values of Curie point $T_C$ and $\Theta_D$ are comparable to those in our systems) such contribution was not found \cite{Ghivelder,Varshney}, but { moderate magnon $BT^{3/2}$ contributions (with $B$ of order of 1 mJ~mol$^{-1}$ K$^{-5/2}$) were found for La$_{1-x}$Sr$_x$MnO$_{3+\delta}$ \cite{Woodfield}}.

The magnon contribution should be qualitatively similar for all our systems with high $T_C$. However, one observes a quite different behavior: the anomalies are large in Cu and Mn systems, whereas they are much weaker for other compounds.
To clarify this problem, we have tried to include the magnon contribution as dominant one for CaCu$_3$Fe$_2$Re$_2$O$_{12}$ where low-frequency rattling modes are not expected, but we did not obtain reasonable fits: the fitted $\gamma$ values even turned out to be negative. Thus we conclude that the magnon contribution is  less important than the lattice anharmonicity and cannot explain the anomalous specific heat behavior.
}

\section{First-principles calculations}

We have already studied rattling in one of the quadrupole perovskites: CuCu$_3$Fe$_2$Re$_2$O$_{12}$~\cite{Pchelkina}. Following the same strategy, we consider in detail another quadruple perovskite based on Mn, which shows a concave-down dependence of $\beta(T)$ according to experimental results presented in Sect.~\ref{exp-data}. MnCu$_3$Fe$_2$Re$_2$O$_{12}$ crystallizes with the same $Pn\overline{3}$ space group with all the ions occupying the same Wickoff positions as, e.g., in CuCu$_3$Fe$_2$Re$_2$O$_{12}$ \cite{Pchelkina}, but has slightly different lattice parameters \cite{MCFRO_Crystal_structure}.

We performed density functional calculations in the Vienna ab initio simulation package (VASP) \cite{Kresse93}, taking into account correlation effects via DFT+U approach in a way, introduced by Liechtenstein et al. \cite{Liechtenstein95}. The Perdew-Burke-Ernzerhof version of exchange correlation potential was used~\cite{Perdew1997}. The cutoff energy for the plane wave basis was set at 500 eV. The convergence criterion was chosen to be 10$^{-6}$ eV, the $6\times6\times6$ k-mesh being used. The calculations were performed using the tetrahedron method with Bl\"ochl corrections. The following values of the onsite Coulomb repulsion and Hund’s exchange parameters were taken, being chosen in \cite{Pchelkina,PchelkinaBiFe,Vorobyova}: $U$(Mn)=3 eV, $J_H$(Mn)=1 eV, $U$(Cu)=7.5 eV, $J_H$(Cu)=1 eV, $U$(Fe)=4.4 eV, $J_H$(Fe)=0.9 eV, $U$(Re)=2 eV, $J_H$(Re)=0.5 eV.

\begin{figure}[t!]
\centering
\includegraphics[width=1\columnwidth]{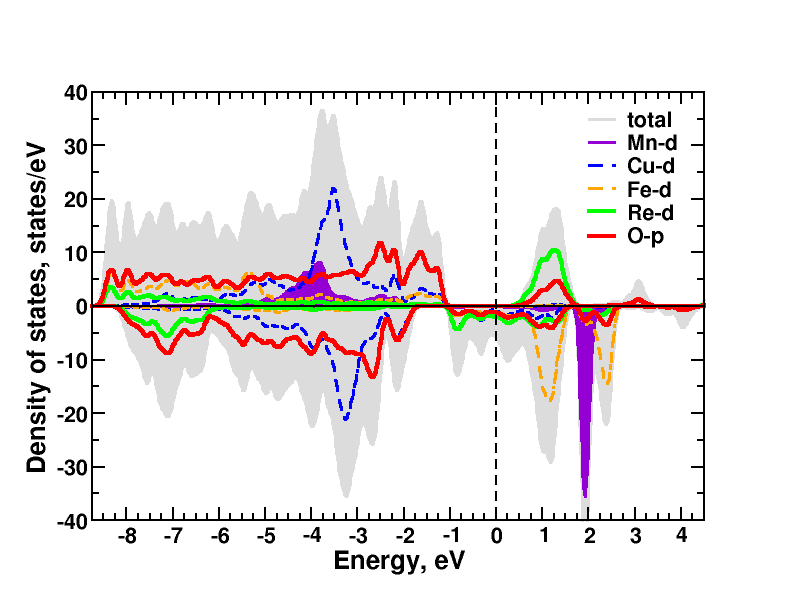}
\caption{\label{DOS-MCFRO} Total and partial density of states calculated in DFT+U for MnCu$_3$Fe$_2$Re$_2$O$_{12}$. Positive (negative) DOS corresponds to spin up (down). The Fermi energy is set to zero.}
\end{figure}

Recent studies of magnetic properties of various quadrupole perovskites suggest that Fe and Re ions are antiferromagnetically (AFM) coupled via superexchange interaction~\cite{Dy,Wang2021}. A stronger (than Cu-Fe) Re-Cu AFM exchange results in the following magnetic structure: Cu($\uparrow$)Fe($\uparrow$)Re($\downarrow$)~\cite{Dy}.  This magnetic configuration was used and  various orientations of Mn magnetic moments were tested. The case with all Mn ions having co-directional to Cu and Fe magnetic moment has the lowest energy. When the neighboring Mn ions have opposite to each other  (i.e., AFM) spin projections, the total energy increases by 56 meV/f.u. and in the case of Mn($\downarrow$)Cu$_3$($\uparrow$)Fe$_2$($\uparrow$)Re$_2$($\downarrow$)O$_{12}$ magnetic configuration, it is also higher by 105 meV/f.u.

The total and partial density of states for the magnetic ground state, i.e., Mn($\uparrow$)Cu$_3$($\uparrow$)Fe$_2$($\uparrow$)Re$_2$($\downarrow$)O$_{12}$, is shown in Fig.~\ref{DOS-MCFRO}. As one can see, the compound is a half-metal, as it is expected for such quadruple perovskites. The resulting magnetic moments are $\mu_{Mn}=4.54 \mu_B$, $\mu_\mathrm{Cu}=0.33 \mu_B$, $\mu_\mathrm{Fe}=4.05 \mu_B$, and $\mu_\mathrm{Re}=-0.78 \mu_B$. Cu-$d$ sub-shell is almost fully occupied, while these are Re-$d$ states, which mostly cross the Fermi level.

In order to simulate rattling, we took the experimental structure and began shifting Mn ions from the center of the surrounding icosahedral oxygen cage. As there are two Mn ions in a primitive cell, both in-phase (when Mn-Mn distance is kept constant, one Mn goes to Fe, the other one to Re) and out-of-phase (when both Mn go either to Fe, or to Re) shifts of Mn ions along [111] direction were considered, see Fig.~\ref{Rattling_types}(a). The resulting energy potential depending on the shift amplitude for this case is given in Fig.~\ref{Potential}(a). As one can see, the largest energy gain, $\sim$30 meV/f.u., is obtained in the case of out-of-phase Mn vibrations to Fe ions.

\begin{figure}[t!]
\centering
\includegraphics[width=1\columnwidth]{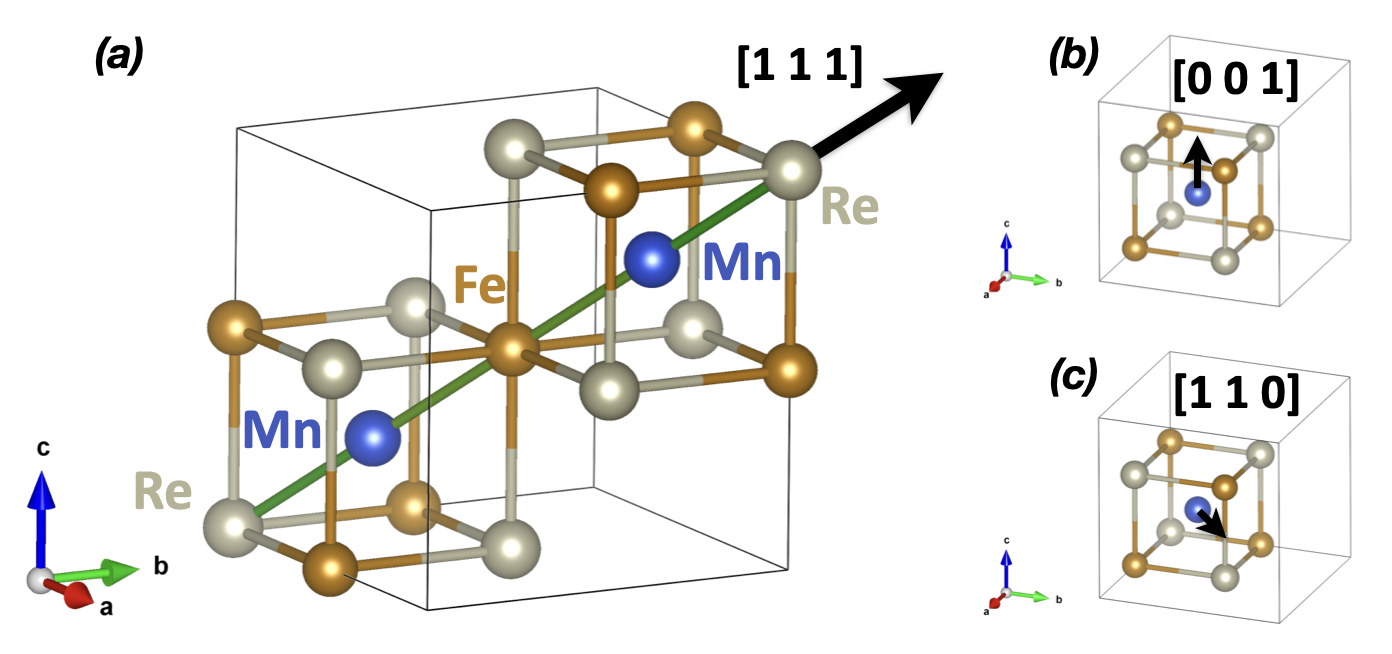}
\caption{\label{Rattling_types} Scheme of the considered rattling distortions in MnCu$_3$Fe$_2$Re$_2$O$_{12}$: (a) demonstrates the whole primitive cell with two Mn ions, (b) and (c) show just half of the cell with Mn shift to the center of Fe$_2$Re$_2$ plane and FeRe edge, correspondingly.}
\end{figure}

\begin{figure}[b!]
\centering
\includegraphics[width=1\columnwidth]{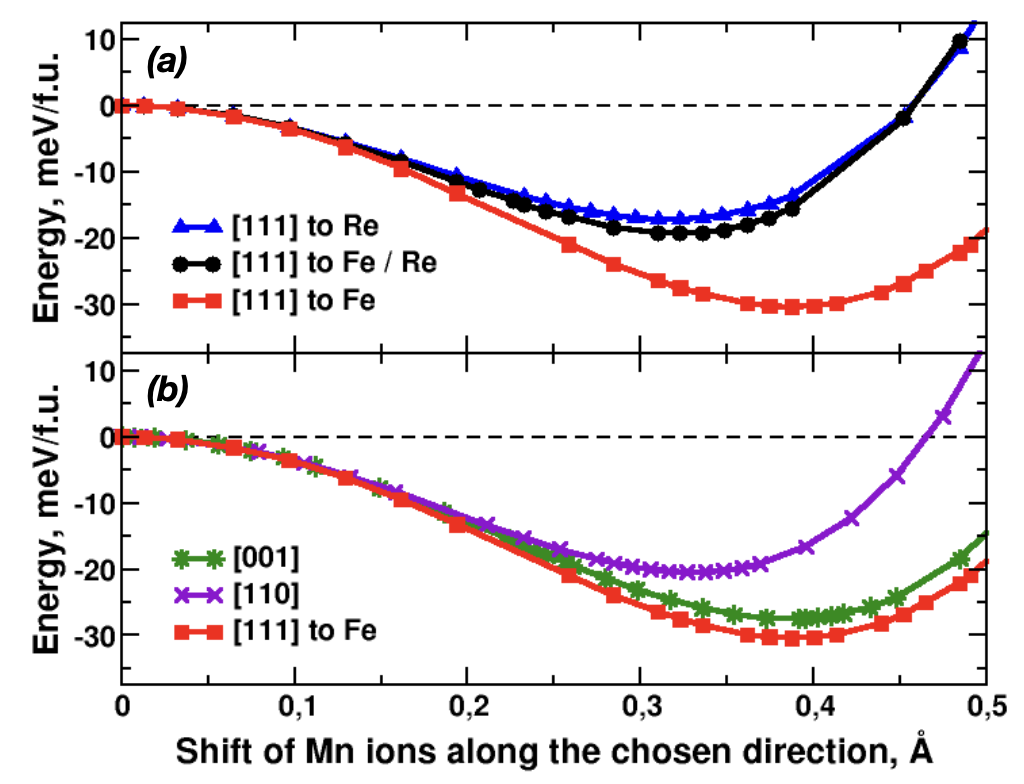}
\caption{\label{Potential} Calculated in DFT+U total energy dependence on the Mn ions displacements (a) along [111] direction, when two different Mn ions in a unit cell are shifted both to Re, one to Re and one to Fe, and both to Fe atoms, and (b) along [001], [110], and [111] directions in MnCu$_3$Fe$_2$Re$_2$O$_{12}$.
}
\end{figure}

Next, other directions of possible rattling distortions, [001] and [110] were checked and it was found that both of them also give an energy gain. Nevertheless, its amplitude, as well as the amplitude of the Mn shift, is lower than for the [111] direction when all Mn ions go to Fe.

These results confirm rattling in MnCu$_3$Fe$_2$Re$_2$O$_{12}$. Moreover, the predicted complicated shape of the energy potential, corresponding to the rattling process, resembles the one obtained for isostructural CuCu$_3$Fe$_2$Re$_2$O$_{12}$ in \cite{Pchelkina}. However, the maximum depth of the potential and the corresponding amplitude of the $A$ ion shift in the compound with Mn is less than for Cu, 30 meV/f.u. $vs$ 55 meV/f.u. and 0.4 \AA~$vs$ 0.55 \AA, correspondingly. Comparison for the crystal structure parameters for these compounds is given in Tab.~\ref{Tab:structures}. As one can see, even though the size of the oxygen cage in MnCu$_3$Fe$_2$Re$_2$O$_{12}$ is larger, Mn$^{2+}$ ion has a larger ionic radius than Cu$^{2+}$ and, therefore, less empty space to vibrate in. In the case of, e.g., CaCu$_3$Fe$_2$Re$_2$O$_{12}$ the sum of ionic radii almost equals two bond lengths between ions and, therefore, there is nearly no space for such Ca shifts.

\begin{table}[t!]
 \begin{tabular}{lccc}
 \hline
 \hline 
$A$-ion                                 & Cu     & Mn     & Ca       \\
\hline
R$_{VI}$($A$), \AA                 & 0.73   & 0.83   & 1.00      \\
R$_{VI}$($A$)+R$_{VI}$(O), \AA     & 2.13   & 2.23   & 2.40      \\
R$_{VIII}$($A$), \AA               &        & 0.96   & 1.12      \\
R$_{VIII}$($A$)+R$_{VIII}$(O), \AA &        & 2.38   & 2.43      \\
R$_{XII}$($A$), \AA                &        &        & 1.34      \\
R$_{XII}$($A$)+R$_{VI}$(O), \AA     &        &        & 2.64      \\
d(A-O), \AA                      & 2.55   & 2.65   & 2.64       \\
 \hline
 \hline
 \end{tabular}
	\caption{Shannon’s ionic radii for Cu$^{2+}$ and Mn$^{2+}$ for large coordination numbers taken from \cite{Shannon1976} and $A$-O bond length in the experimental crystal structure for $A$Cu$_3$Fe$_2$Re$_2$O$_{12}$ ($A$=Cu, Mn, Ca). Subscripts refer to the coordination number. For oxygen ion, Shannon's ionic radius is $R_{VI}$(O)= 1.40 \AA~and $R_{VIII}$(O)=1.43 \AA~ \cite{Shannon1976}.}
		\label{Tab:structures}
\end{table}

A similar trend is observed in the experimental values of the atomic thermal displacement parameters for the the V-containing compounds CuCu$_3$V$_4$O$_{12}$, MnCu$_3$V$_4$O$_{12}$, and CaCu$_3$V$_4$O$_{12}$~\cite{Akizuki2015}. This tendency is attributed to the decrease in the empty space, that is, the difference between the cage size and the ionic radius of the $A$-site cation, along this sequence. Notably, the strongest anomalous temperature dependence of specific heat (low $\Theta_E$) was experimentally observed for Mn and Cu (see Sec.~\ref{exp-data}). For both materials, our DFT+U calculations clearly predict rattling of $A$ ions. 

{ According to the powder X-ray diffraction data \cite{Cu,MCFRO_Crystal_structure}, unusually large $U_{iso}$ observed for the A-site Cu (Cu1) in CuCu$_3$Fe$_2$Re$_2$O$_{12}$  and Mn in MnCu$_3$Fe$_2$Re$_2$O$_{12}$  can be attributed to either static disorder or enhanced thermal vibrations. To evaluate the possibility of static disorder, we first tested for $A$-site Cu1 and Mn deficiencies, but the refinements showed no significant deviation from full occupancy ($g$(Cu1/Mn) = 1). We then explored split-site models involving Cu1 and Mn displacements, but none of these models led to improved R-factors. These results effectively rule out displacive disorder at the $A$-site, supporting the interpretation that the large $U_{iso}$ values arise primarily from vibrations of the Cu1 and Mn atoms. The dynamic Cu1/Mn oscillation also has close relationship with an unusual behavior of specific heat described above.}

\section{Possible mechanisms of rattling}
One of the key questions is a mechanism of  rattling. The displacement of $A$ ions away from the high-symmetry position in rattling reminds the Jahn-Teller (JT) effect, where the  energy is gained by removing orbital degeneracy due to lowering the  symmetry~\cite{Ceulemans2022,Khomskii-book}.  In our case, these are rare-earth or transition-metal ions with completely filled $d$-shell, which typically occupy positions available for rattling ($A$ sites). Therefore, one cannot lift any orbital degeneracy in this way, but the very similar physics appears in pseudo-Jahn-Teller effect (sometimes called second-order JT effect): the total energy of a system can be significantly lowered by, e.g., the mixing of occupied and empty states (both are non-degenerate) and the distortion of a local surrounding \cite{Opik1957,Fulton1961,Bersuker-book}.

This effect successfully explains the formation of the spontaneous electric polarization due to the distortions breaking the inversion symmetry in many ferroelectrics. For example, in famous BaTiO$_3$ these distortions are due to mixing of completely filled O $2p$ states and  empty Ti $3d$ states. The vibronic coupling  additionally lowers the total energy due to hybridization between O $2p$ and Ti $3d$ states~\cite{Bersuker1966}. 
\begin{figure}[t!]
\centering
\includegraphics[width=1\columnwidth]{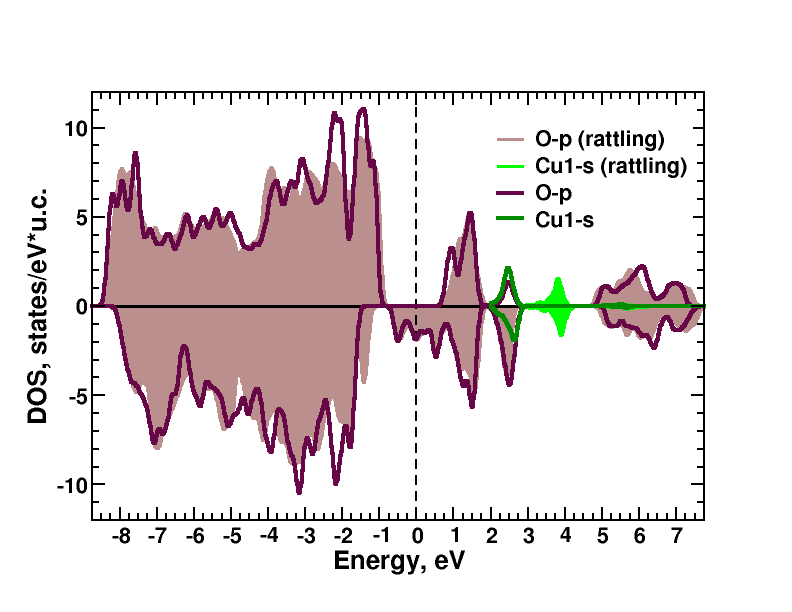}
\caption{\label{DOS} Partial density of states for O-$p$ and Cu1-$s$ orbitals in CuCu$_3$Fe$_2$Re$_2$O$_{12}$ calculated in DFT+U. Positive (negative) DOS corresponds to spin up (down). DOS's in the experimental (undistorted) crystal structure are given in dark brown and green lines. DOS's corresponding to the rattling along [111] direction with the amplitude obtained in \cite{Pchelkina} are given in light brown and light green. The Fermi energy is set to zero. }
\end{figure}

In our case, this effect can be, e.g., mixing of the  empty $s$ states of a transition metal at the $A$ sites and occupied O-$2p$ orbitals having different parities. Rattling distortion shifts $s$ states of the TM to the higher energies, but at the same time changes electronic structure of the occupied states. Competition between the energy gain due to this modification and the loss in the elastic energy determines the equilibrium crystal structure. In Fig.~\ref{DOS} we present density of electronic states plot of the highly-symmetric cubic phase and the distorted by rattling structure for CuCu$_3$Fe$_2$Re$_2$O$_{12}$. One can clearly see that, indeed, rattling pushes empty Cu1 $s$ states up in energy. Calculations of the center of gravity of occupied part of O $2p$ band demonstrate that it shifts downwards in the distorted phase lowering the total energy, which is consistent with the pseudo-Jahn-Teller scenario.

While it is tempting to attribute rattling to the pseudo-Jahn-Teller effect, other mechanisms for this phenomenon may exist. A detailed investigation and microscopic calculations of both chemical bonding and  other contributions to the total energies is necessary to develop a comprehensive and realistic theory of rattling.

\section{Conclusions} 

Our experimental results definitely demonstrate anomalous temperature dependence of the specific heat, which give evidence of anharmonic effects in the lattice subsystem of quadruple perovskites CuCu$_3$Fe$_2$Re$_2$O$_{12}$  and MnCu$_3$Fe$_2$Re$_2$O$_{12}$. 
Although similar effects have been  observed earlier in a number of systems mentioned in the Introduction, the anomalies in quadruple perovskites studied are highly pronounced (characteristic temperature scales given by the fitted Einstein temperatures are lower).   { Refining of X-ray diffraction data \cite{Cu,MCFRO_Crystal_structure} demonstrates very large isotropic displacement parameters for Cu and Mn ions sitting in icosahedra in CuCu$_3$Fe$_2$Re$_2$O$_{12}$  and MnCu$_3$Fe$_2$Re$_2$O$_{12}$ suggesting rattling.}

Together with theoretical DFT+U calculations, our specific heat data support the existence of rattling in the  perovskites $A$Cu$_3$Fe$_2$Re$_2$O$_{12}$, especially for $A$ = Cu and Mn. In the case of rare-earth metals occupying the $A$ sites some temperature dependence of the characteristic function $\beta(T) = (C_p - \gamma T)/T^3$ is also observed, but further studies are necessary. Theoretically, these systems are more challenging to analyze because of many-body effects arising from both a larger Coulomb interaction and stronger spin-orbit coupling. 

Since low-energy lattice vibrations serve as efficient scattering centers, one could expect anomalous suppression of the thermal conductivity  in rattling systems, which can be observed experimentally  (cf. \cite{Bauer,Baumbach}). { Thus, quadruple perovskites and related systems are promising candidates for thermoelectric devices.
To conclude, investigation of thermodynamic properties of quadruple perovskites is of great physical interest and provides a potential for practical applications.}

\section{Acknowledgments}

SVS is grateful to L. Chibotaru and F. Teminikov for fruitful discussions. 
 Y. Long was supported by the National Key R\&D Program of China (Grant No. 2021YFA1400300), and the National Natural Science Foundation of China (Grant Nos. 12425403, 12261131499).
The first-principle calculations were supported by the Russian Science Foundation via project RSF 23-42-00069. 
The analysis of specific heat data in quadruple perovskites was performed   within the framework of the state assignment of the Ministry of Science and Higher Education of the Russian Federation for the IMP UB RAS.

\end{document}